%% file: apj_shallow.tex
\documentclass[manuscript]{aastex61}

\usepackage{amsmath}
\usepackage[utf8]{inputenc}
\usepackage[T1]{fontenc}
\usepackage{newtxtext, newtxmath}

\usepackage{rotating}  

\usepackage{savesym}
\savesymbol{tablenum}
\usepackage{siunitx}
\restoresymbol{SIX}{tablenum}

\newcommand{\vect}[1]{\boldsymbol{#1}}
\newcommand{\cvect}[1]{\boldsymbol{\mathrm{#1}}}

\newcommand{\dd}[2]{\frac{\partial #1}{\partial #2}}
\newcommand{\diff}[2]{\frac{\d{} #1}{\d #2}}

\renewcommand*\d{\mathop{}\!\mathrm{d}}

\newcommand\grav{g}
\newcommand\tfric{\tau_\C{drag}}
\newcommand\trad{\tau_\C{rad}}
\newcommand\twave{\tau_{\C{wave}}}
\newcommand\heq{h_\C{eq}}

\newcommand{\C}[1]{\mathrm{#1}}  
\newcommand\Ro{\mathrm{Ro}}

\begin{document}
	\title{The Thermal Phase Curve Offset on Tidally- and Non-Tidally-Locked Exoplanets: a Shallow Water Model}
	\author{James Penn}
	\affil{University of Exeter}
	\email{jp492@exeter.ac.uk}
	\author{Geoffrey K Vallis}
	\affil{University of Exeter}
	\email{g.vallis@exeter.ac.uk}
	\begin{abstract}
		\input{abstract}
	\end{abstract}
	\input{introduction}
	\input{model}
	\input{results}
	\input{discussion}
	\bibliography{full_biblio}
\end{document}

%% file: abstract.tex

Using a shallow water model with time-dependent forcing we show that the peak of an exoplanet thermal phase curve is, in general, offset from secondary eclipse when the planet is rotating.
That is, the planetary hot-spot is offset from the point of maximal heating (the substellar point) and may lead or lag the forcing; the extent and sign of the offset is a function of both the rotation rate and orbital period of the planet.
We also find that  the system reaches a steady-state in the reference frame of the moving forcing.
The model is an extension of the well studied Matsuno-Gill model into a full spherical geometry and with a planetary-scale translating forcing representing the insolation received on an exoplanet from a host star.

The speed of the gravity waves in the model is shown to be a key metric in evaluating the phase curve offset.
If the velocity of the substellar point (relative to the planet's surface)  exceeds that of the gravity waves then the hotspot will lag the substellar point, as might be expected by consideration of forced gravity wave dynamics.
However, when the substellar point is moving slower than the internal wavespeed of the system the hottest point can lead the passage of the forcing.
We provide an interpretation of this result by consideration of the Rossby and Kelvin wave dynamics as well as, in the very slowly rotating case,  a one-dimensional model that yields an analytic solution.
Finally, we consider the inverse problem of constraining planetary rotation rate from an observed phase curve.

%% file: introduction.tex

\section{Introduction} 
\label{sec:introduction}

In 2007 the first thermal map of an exoplanet was obtained from transit recordings of hot Jupiter HD189733b \citep{Knutson:2007bl}, showing that the hottest point on the surface of the planet was not at the substellar point, but offset eastward. These observations were consistent with previous GCM studies of tidally-locked `hot Jupiters', in which equatorial superrotating jets in the atmosphere provided zonally-asymmetric heat transport from day to night side \citep{Showman:2002ez}.
More recently, the super-Jupiter 2M1207b became the first planet for which a rotation rate has been constrained using direct-imaging from an intense study with Hubble \citep{Zhou:2016gc}. While direct-imaging is the optimal method for measuring the parameters of an exoplanet, this is impractical for many small and close-in planets (2M1207b is separated from its star by a distance of \SI{41.2}{AU}) where the brightness of the host star and variability in stellar output preclude direct-imaging.

Current resolution limits have restricted transit detection to large, close-in, hot Jupiters across G-class stars (like our own Sun); smaller rocky planets here remain undetectable. Still, there has been interest in using the same techniques to observe low mass, cooler stars -- dwarfs or ``ultra-dwarfs'', around which close-in rocky exoplanets of approximately one Earth mass can be resolved \citep{Gillon:2016hl}.
With effective temperatures of 200 -- \SI{400}{K} such planets may well be capable of supporting atmospheres similar to that of Earth or Venus; characterising the atmospheric dynamics prevailing on the planet will help deduce whether a planet could be considered ``habitable'' or not, and resolving the thermal phase curve of this class of planets from transit observations gives us insight into the prevailing conditions in the thermally emitting layer of the planetary atmosphere.

By measuring the total infrared emission of a planetary system at all points in a planet's orbit, and then subtracting the emission of the star alone, the thermal phase curve of the planet can be constructed. When a planet passes in front of its host star, the primary eclipse is observed and a sharp reduction in light intensity is recorded as the planet obscures part of the star. As the observer is receiving both light from the star and light from the planet, there is another smaller amplitude secondary eclipse as the planet goes behind the star and planetary emission and reflected light are blocked by the star.
The amplitude of the curve can tell allow us to measure day to night temperature difference, the peak of the phase curve relative to the secondary eclipse gives us information of the hottest face of the planet.  In general this is not the same as the face receiving the most stellar insolation; for example, the phase curve of 55 Cancri e, a hot super-Earth in a close orbit, shows $41\pm12$ degree eastward offset of the hotspot from the substellar point \citep{Gillon:2016kg}.
To help understand such phase curves is often assumed that, because of the strong tidal forces exerted by the host star, close-in exoplanets should be tidally-locked to the host star, and much research has naturally focussed on their properties. Thus,  an analytic theory for the day-night temperature difference on tidally-locked hot Jupiters has been developed \citep{PerezBecker:2013ik} and, developing the ideas of hot Jupiter heat distribution further, \cite{Zhang:2016te} (their appendix B) provided a theoretical model for the thermal phase curve shift, demonstrating with the use of a GCM that for planets with a superrotating equatorial jet, the hotspot offset can be parameterised by the ratio of radiative to zonal-jet advective timescales.

However, thermal tides in an atmosphere can, in some cases, be strong enough to force a planet out of a synchronous rotation; this effect is seen in the case of Venus's slow retrograde rotation \citep{Ingersoll:1978ht,Dobrovolskis:1980cf}. Furthermore, recent GCM simulations \citep{leconte:2015kk} demonstrate that while tidal frictional forces slow the rotation rate of a planet, as orbital radius increases a thin atmosphere of merely 1 bar is sufficient to generate thermal tides strong enough to maintain a planet out of synchronous rotation. Also, although many planets may be evolving toward a tidally-locked state they may not yet be there. For all these reasons, the assumption that a rocky terrestrial planet closely orbiting a low-mass star is tidally-locked may not always be valid, and as our detection methods improve and we resolve planets in wider orbits we may discover planets that escape tidal-locking altogether, like Earth. Certainly, for large orbital radius the rotational rate of the planet may be essentially independent of the orbital period and primarily influenced by other factors such as the way the planet was formed, satellites or other nearby planets and stars.

In any planet that is not tidally locked the solar forcing will appear, in the frame of reference of the planet's surface, to be moving, and this problem has been much less extensively studied than the tidally-locked counterpart, albeit with some  exceptions -- notably the ``moving-flame'' rotating-tank laboratory experiments of \citet{Schubert:1969fe} and the linearised shallow water study of  \citet{Kato:1994wb}.  In the latter, by varying the Lamb parameter $\epsilon = 4a^2\Omega^2/gH$, the strength of drag forces and the velocity of the moving forcing, the authors classified the steady-state solutions into four categories; direct circulation between the day and night hemispheres, ``Gill pattern'' circulations, zonally symmetric flow and finally a mode of resonant inertio-gravity waves. To a large extent, the roles of drag and of rotation were found to be interchangeable in the linear model; the same flow patterns emerging under fast rotation - high drag as for slow rotation - low drag. The circulation regime observed in the shallow water system with mobile forcing has a stronger dependence on the frictional timescale than the radiative timescale;  the simplifying assumption chosen by \cite{Gill:1980bma} and \cite{Matsuno:1966wt} - that Rayleigh friction and Newtonian cooling timescales are equal - was shown to be appropriate when the timescales are of the same orders of magnitude \citep{Kato:1997ti}.

In this paper we further examine the consequences of non-tidal locking, and we specifically address the question of how relaxing the assumption of tidal-locking on a terrestrial exoplanet affects the observed phase curve.
As with a number of earlier studies \citep[e.g.,][as well as those cited above]{Cho:2003dy, Showman:2011cb} we use a shallow water model,  allowing us to unpack cause and effect in a way that is not possible in more complex GCMs with parameterised physics.   Specifically, we adopt a similar model to \cite{Kato:1997ti}, extending it to an examination of the full non-linear equations on the sphere and a larger parameter range of forcing velocity, moving both in a prograde and retrograde direction, using the Matsuno-Gill approximation of equal timescales for both radiative cooling and fictional drag.
In the case of a slowly moving forcing and slowly rotating planet, the results are interpreted using a one-dimensional linearised model along the equator, where we derive an analytic form for the amplitude and offset of the phase curve. The outline of the paper is as follows. In Section \ref{sec:model} we introduce the model itself, the parameters we use and the method of solution. In Section \ref{sec:results} we describe the results, in Section \ref{sec:discussion} we discuss and interpret those results, and in Section \ref{sec:conclusions} we give some concluding remarks.


%% file: model.tex

\section{Model}
\label{sec:model}

Starting with the primitive equations and expanding the vertical structure into normal modes of height, we can derive a set of horizontal shallow water equations with an associated equivalent depth $H$ (e.g. \citealt{Vallis:2006wt,Schubert:2006ii}).
Here we examine the first baroclinic mode, that is, the first vertical mode deviating from the barotropic mean.

\begin{figure}[tb]
  \centering
  \includegraphics[width=0.8\textwidth]{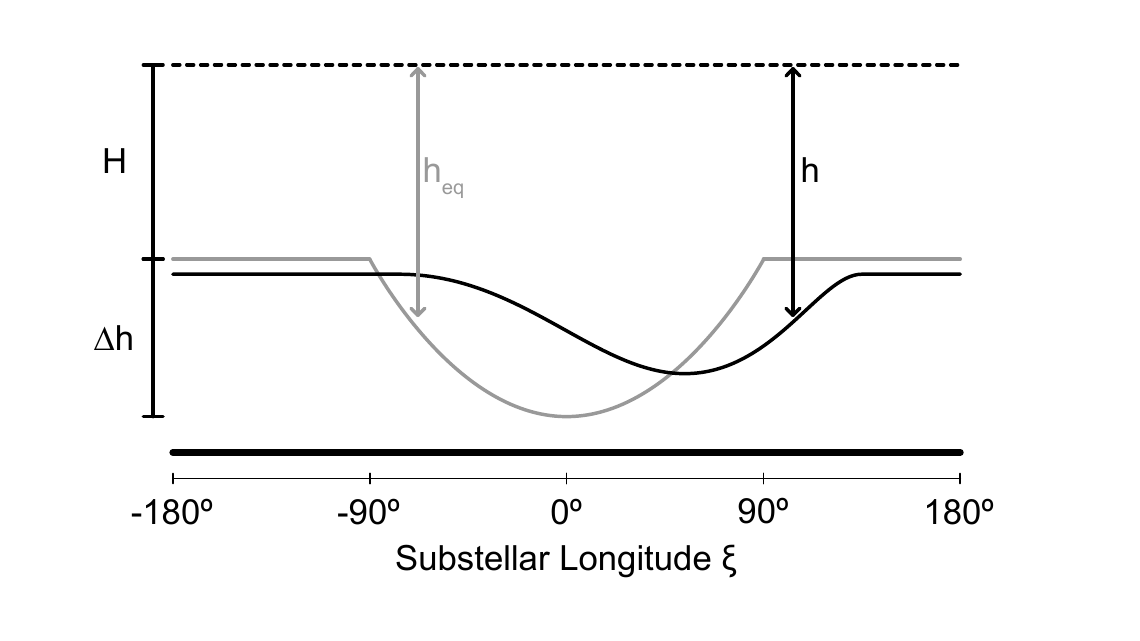}
  \caption{The equilibrium profile $\heq$ represents the heating effect as a thickening of the geopotential of the upper-atmosphere.
  Where the stellar insolation irradiates the day side of the planet, the geopotential $gh$ is forced towards a deeper equilibrium depth.
  The rate at which the geopotential is forced towards the equilibrium profile is determined by the radiative cooling timescale $\trad$.}
  \label{fig:model}
\end{figure}

We use a single layer shallow water model
\begin{align}
  &\dd{\vect u}{t} + \vect{u} \cdot \nabla \vect{u} + \vect{f} \times \vect u = - \grav \nabla h - \frac{\vect{u}}{\tfric}, \label{eqn:u} \\
  &\dd{h}{t} + \nabla \cdot (\vect u h) = \frac{h_{\C{eq}} - h}{\trad}, \label{eqn:h}
\end{align}
where $\vect{u}$ is the horizontal velocity, $\vect{f}= 2 \Omega \sin \phi \, \hat{\cvect{k}}$ is the Coriolis vector, $\grav$ the gravitational acceleration and $\tfric$ and $\trad$ are Rayleigh frictional drag and Newtonian radiative cooling timescales respectively.
The equations are considered on the surface of a sphere, with a spherical coordinate system of latitude, $\phi$, and longitude, $\lambda$.
The height of the fluid layer in this model is a proxy to temperature - a thickening of the layer corresponding to a higher temperature in the upper troposphere and thus higher emission temperature (Figure~\ref{fig:model}).

\begin{figure}[tb]
	\centering
	\includegraphics[width=0.6\textwidth]{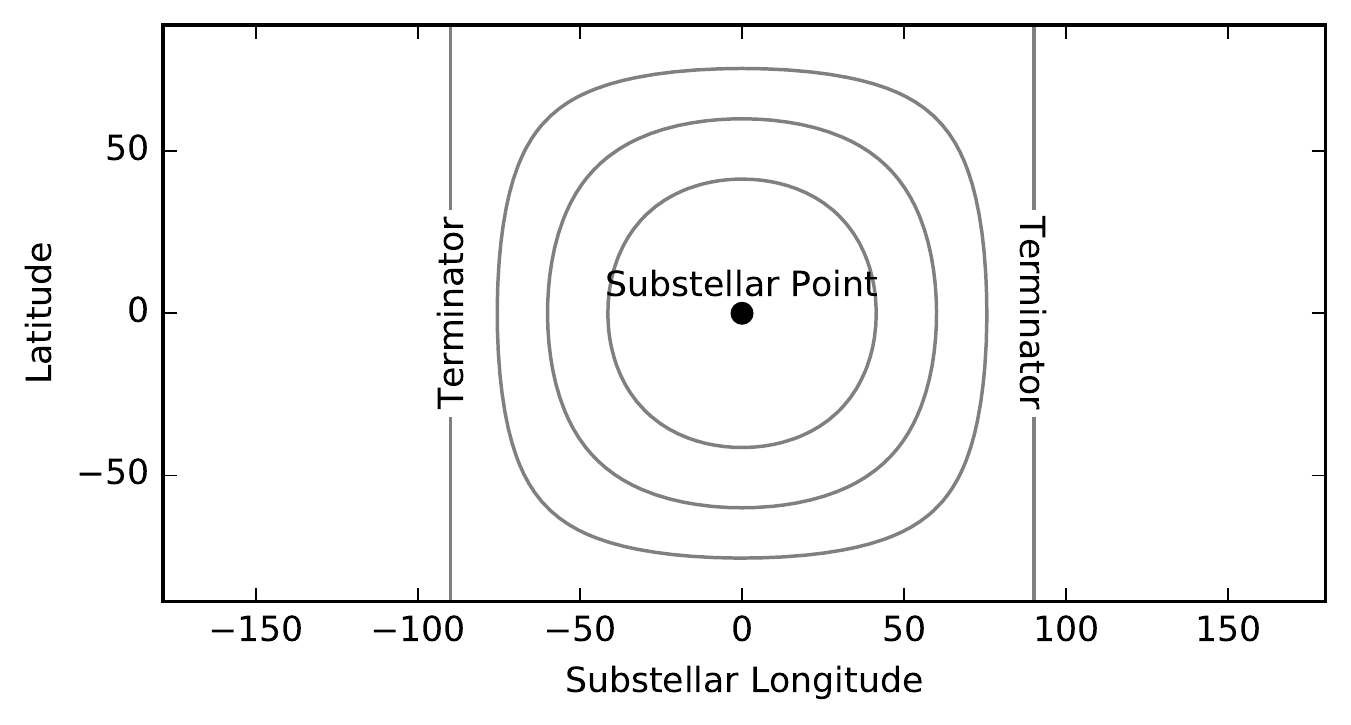}
	\caption{Contours of the equilibrium profile $\heq$.  It is stationary in latitude $\phi$ and substellar longitude $\xi$ with chines at $\pm \pi/2$ corresponding to the dawn and dusk terminators.}
	\label{fig:heq}
\end{figure}

The equations are forced by a relaxation to an equilibrium profile $h_\C{eq}$, the height field corresponding to radiative equilibrium
\begin{equation}
  h_\C{eq}(\lambda, \phi, t) = H + \Delta h \cos \phi \max(\cos(\lambda - \lambda_0, 0)),
\end{equation}
parameterised by the substellar longitude $\lambda_0$ (Figure~\ref{fig:heq}).
$H$ is the reference fluid height in the absence of stellar forcing.
In this choice of forcing we are implying zero obliquity, eccentricity and procession.
To constrain the problem to a manageable parameter space, in this study we set the values of $\trad = \tfric = \tau$.
As has been discussed elsewhere \citep{Showman:2011cb, Kato:1997ti}, the true ratio of these timescales in an exoplanet atmosphere may be significantly different; future work will be to extend the study to varying these parameters independently.

On a tidally-locked exoplanet, the substellar point $\lambda_0$ remains fixed in longitude.
However, for asynchronously rotating planet with zero obliquity and eccentricity the substellar point will travel along the equator with constant velocity, inducing a regular planetary diurnal cycle.
The diurnal period on a planet is given by the difference between rotation rate and orbital rate
\begin{equation} \label{eqn:tsol}
  \C{P_{sol}} = \frac{2 \pi}{\Gamma - \Omega},
\end{equation}
where $\Gamma$ is the orbital rate of the planet, $\Gamma = 2 \pi/\C{P_{orb}}$ for orbital period $\C{P_{orb}}$.
The length of a stellar day on a planet is then $\C{T_{sol}} = |\C{P_{sol}}|$.
The sign of $\C{P_{sol}}$ is important for determining the longitude of the substellar point and its direction of travel across the planet.
At time $t$, the substellar point is located at longitude
\begin{equation} \label{eqn:sublon_tsol}
  \lambda_0(t) = 2 \pi  \frac{t}{\C{P_{sol}}} = (\Gamma - \Omega)t.
\end{equation}
For Earth, which has positive  $\Omega_\Earth > \Gamma_\Earth > 0$ (anti-clockwise rotation/orbit when viewed from the North pole) the subsolar point tracks east-west across the surface of the planet, i.e. $\d\lambda_0 /\d t < 0$.

While the ratio of orbital to planetary rotation rates determines whether a planet is in synchronous or asynchronous rotation, from an atmospheric dynamics perspective and in the context of our shallow water model, we are concerned only with the manifestation of this rate differential: the length of the diurnal cycle and the velocity of the substellar point as it traverses the planetary surface.

The non-dispersive Kelvin wave speed $c = \sqrt{gH}$ determines the maximal information velocity in the shallow water equations and provides a natural velocity scaling for the system.
Let $x_0$ be the location along the equator, in \si{m}, of the substellar point from the origin.
Then $x_0 = a \lambda_0$, where $a$ is the radius of the planet, here fixed at $a = \SI{6371}{km}$.
Substituting $x_0$ into (\ref{eqn:sublon_tsol}) and taking the time derivative we obtain the velocity of the substellar point (denoted $s$)
\begin{equation}
  s = \diff{x_0}{t} = a (\Gamma - \Omega) \equiv \alpha c. \label{eqn:substellar_vel}
\end{equation}
We define the non-dimensional parameter $\alpha = s/c$ which will be varied in our numerical simulations to set the velocity of the substellar point in terms of the wavespeed of the shallow water layer.
In this parameter scheme, $\alpha = 0$ for a tidally-locked planet.
When $\alpha < 0$ the substellar point moves retrograde (as on Earth) and when $\alpha > 0$ it is prograde, moving west to east in the direction of planetary rotation.

Lastly we introduce a new longitudinal coordinate with origin at the substellar point $\xi = \lambda - (\alpha c/a) t$ so that the equations solved numerically are
\begin{align}
  \dd{\vect{u}}{t} + \vect u \cdot \nabla \vect u + 2 \Omega \sin \phi \hat{\cvect{k}} \times \vect{u} &= - \grav \nabla h - \frac{\vect{u}}{\tau}, \label{eqn:model_u} \\
  \dd{h}{t} + \nabla \cdot (\vect{u} h) &= \frac{h_{\C{eq}} - h}{\tau}, \label{eqn:model_h} \\
  h_{\C{eq}} &= \begin{cases}
	H  + \Delta h \cos\phi \cos\xi & \cos\xi \ge 0 \\
	H  & \cos\xi < 0, \\
	\end{cases} \label{eqn:model_heq} \\
  \xi &= \lambda - \frac{\alpha c}{a} t. \label{eqn:xi}
\end{align}
The model is parameterised by the night-side relaxation height $H$, day-side forcing scale $\Delta h$, rotation rate $\Omega$, substellar velocity $\alpha c$ and frictional timescale $\tau$.

Experiments were performed using a small ($\Delta h = 0.1H$) and large ($\Delta h = H$) scale forcing.
The final steady-state solutions, normalised to the scale of $\Delta h$, are quantitatively similar for both large and small values of $\Delta h$;
all results presented below are from the $\Delta h = 0.1 H$ experiments.

The equations (\ref{eqn:model_u}) - (\ref{eqn:xi}) were integrated numerically using a pseudospectral core at T85 ($128 \times 256$ in latitude-longitude) resolution.
To maintain numerical stability a weak 4th order hyperdiffusivity term is included in both the vorticity and divergence prognostic equations.

%% file: results.tex

\section{Results} 
\label{sec:results}

A parameter sweep varying planetary rotation rate, $\Omega$, from \SI{1e7}{s^{-1}} to \SI{5e-4}{s^{-1}} and $\alpha$ from -2 to 2 was performed.
For a given value of $\Omega$ and $\alpha$, the numerical model described above was initialised in a quiescent state and integrated forward in time until the solution converged to a steady-state in the $(\xi, \phi)$ reference frame.  We also define a \emph{planetary Rossby number} $\Ro = \sqrt{gH}/ (\Omega a)$, which is the ratio of the deformation radius $L_d = \Omega/\sqrt{gH}$ to planetary radius $a$.
Associated with this we define a timescale of global wave propagation $\tau_\C{wave} = a / \sqrt{gH}$.

\begin{sidewaysfigure}
	\centering
			\includegraphics[width=\textwidth]{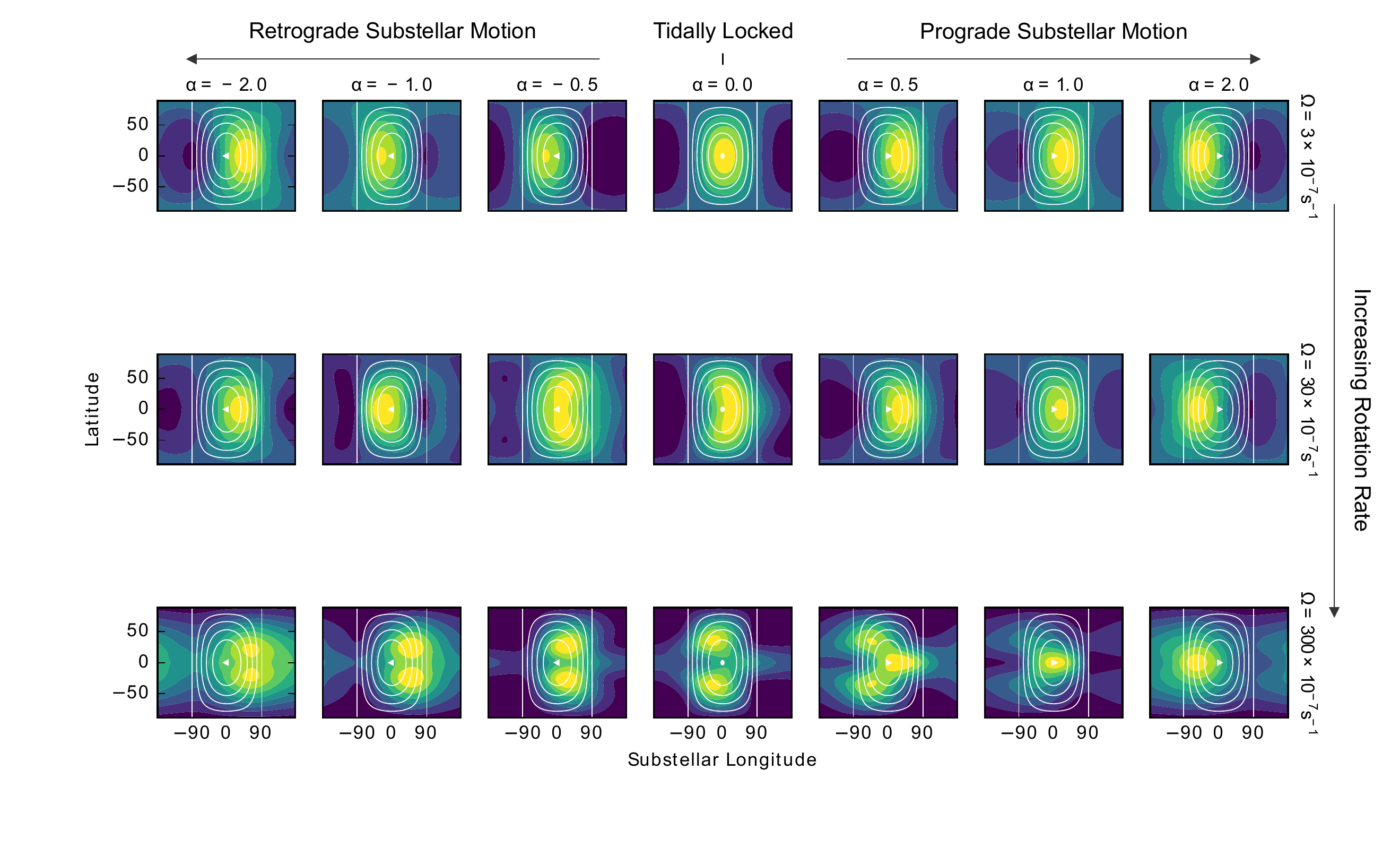}
		\caption{Steady-state solutions for varying $\Omega$ and $\alpha$. Geopotential is shown in filled coloured contours, rescaled for maximal contrast (each colour map has a different scale).
		White contour lines indicate the shape and position of the forcing, $\heq$; the location and direction of the substellar point at ($\xi=0$, $\phi=0$) is denoted by a white arrowhead.
		Columns show experiments with increasing values of $\alpha$, varying subsolar velocity from $-2c$ ($\alpha = -2$) to $2c$ ($\alpha = 2$).
		Rows from top to bottom demonstrate increasing planetary rotation rate for three representative cases of slow, intermediate and fast rotation.  $gH=\SI{1000}{m^{2}s^{-2}}$, $\tau = \SI{5}{days}$}
		\label{fig:omega_spin_space}
\end{sidewaysfigure}

Figure~\ref{fig:omega_spin_space} shows steady-states obtained by varying substellar velocity and planetary rotation rate.
From top to bottom, three representative cases of $\Omega$ are given in the rows of slow ($\Ro \simeq 15$), medium ($\Ro \simeq 1.5$) and fast ($Ro \simeq 0.15$) planetary rotation.
The central column, $\alpha = 0$, corresponds to a tidally-locked planet with increasing rotation rate.
Qualitatively, the tidally-locked steady-state solutions in the central column appear similar in structure to both the original work of \cite{Gill:1980bma} (in the fast rotating case) as well as previously published tidally-locked shallow water \citep{Showman:2010hk} and hydrostatic three-dimensional GCMs (for example \cite{Komacek:2016ij}).

For the slowly rotating system ($\Omega = \SI{3e-7}{\per\second}$, top row) planetary Rossby number is large and substellar velocity small, drag forces balance pressure gradient and geopotential largely relaxes to the forcing profile.
When the substellar point is moving slowly in either a prograde or a retrograde direction ($|\alpha| < 1$), the maximum height perturbation can be seen to be leading ahead of the substellar point.
In the rapid rotating case ($\Omega = \SI{300e-7}{\per\second}$, bottom row) Rossby number is small and we observe a large influence of the differential rotation rate between equator and midlatitudes; the geopotential anomalies are centred away from the equator and have character similar to the classic ``Matsuno-Gill'' pattern described by \cite{Matsuno:1966wt} and \cite{Gill:1980bma}.

In the extreme left and right columns the substellar point is moving faster than the gravity wave speed; the geopotential response can be seen to be lagging behind the motion of the forcing, the peak at stasis downstream of the substellar point.
When the planet is rotating slower than a critical rate (top and middle rows), the hottest point on the planet remains on the equator and ahead of the substellar point, whether it is moving prograde or retrograde.
At the critical rotation rate the character of the solution transitions into the Matsuno-Gill regime: the hottest point splits into two and moves into a meridionally symmetric pattern in the tropics.
These are Rossby gyres, equatorially trapped Rossby waves, propagating westward and damped by the radiative and frictional forces introduced over timescale $\tau$.

\subsection{Phase curves} 
\label{sub:phase_curves}

When observing a spherical planet from a distance we see only a single hemisphere at any one time.
The total thermal emission from the hemisphere is received as a single point reading, as a planet orbits its host star we can observe different hemispheres and record different temperatures -- once done for all angles in the orbit the thermal phase curve can be calculated.
With fluid height representing atmosphere geopotential thickness -- a proxy to temperature -- we can calculate an ``emission phase curve'' from the shallow water model by performing disc integrals of the height field
\begin{equation}
	I(\delta) = \int_{\delta - \pi/2}^{\delta + \pi/2} \int_{-\pi/2}^{\pi/2} a^2 h(\lambda, \phi) \cos\lambda \cos^2\phi \d\phi  \d\lambda,
\end{equation}
where $\delta$ is the observational zenith longitude.  The $\cos \lambda \cos \phi$ factor comes from the projection of the curved surface of the planet onto a flat observational disc -- emission received is proportional to the distance from the centre of the disc.

The phase curve is normalised over the range of the equilibrium profile, $\heq$, by calculating the day and night side temperatures that the model is being forced towards.
Considering the the day ($\cos \xi \ge 0$) and night ($\cos \xi < 0$) branches of (\ref{eqn:model_heq}), we calculate bounds on the equilibrium phase curve
\begin{align}
	I_\C{\mathrm{night, eq}} &= \int_{-\pi/2}^{\pi/2} \int_{-\pi/2}^{\pi/2} a^2 H \cos^2 \phi \cos \lambda \d \phi \d \lambda = \pi a^2 H, \\
	I_\C{\mathrm{day, eq}} &= \int_{-\pi/2}^{\pi/2} \int_{-\pi/2}^{\pi/2} a^2 (H+\Delta h \cos \phi \cos \lambda) \cos^2 \phi \cos \lambda \d \phi \d \lambda = I_\C{\mathrm{night, eq}} + \frac{2}{3} \pi a^2 \Delta h.
\end{align}
The normalised phase curve function is then given by
\begin{align}
	\hat I(\delta) &= \frac{\int_{\delta - \pi/2}^{\delta + \pi/2} \int_{-\pi/2}^{\pi/2} a^2 h(\xi, \phi) \cos\xi \cos^2\phi \d\phi  \d\xi - I_\C{\mathrm{night, eq}}}{(I_\C{\mathrm{day, eq}} - I_\C{\mathrm{night, eq}})} \\
	&=  \frac{3}{2} \frac{\int_{\delta - \pi/2}^{\delta + \pi/2} \int_{-\pi/2}^{\pi/2} h(\xi, \phi) \cos\xi \cos^2\phi \d\phi  \d\xi - \pi H}{\pi \Delta h},
\end{align}
where observational zenith longitude, $\delta$, is relative to the substellar point.

Normalised phase curves were calculated for the steady-state solutions for the simulations varying both $\Omega$ and $\alpha$.
\begin{figure}[tb]
	\centering
	\includegraphics[width=\textwidth]{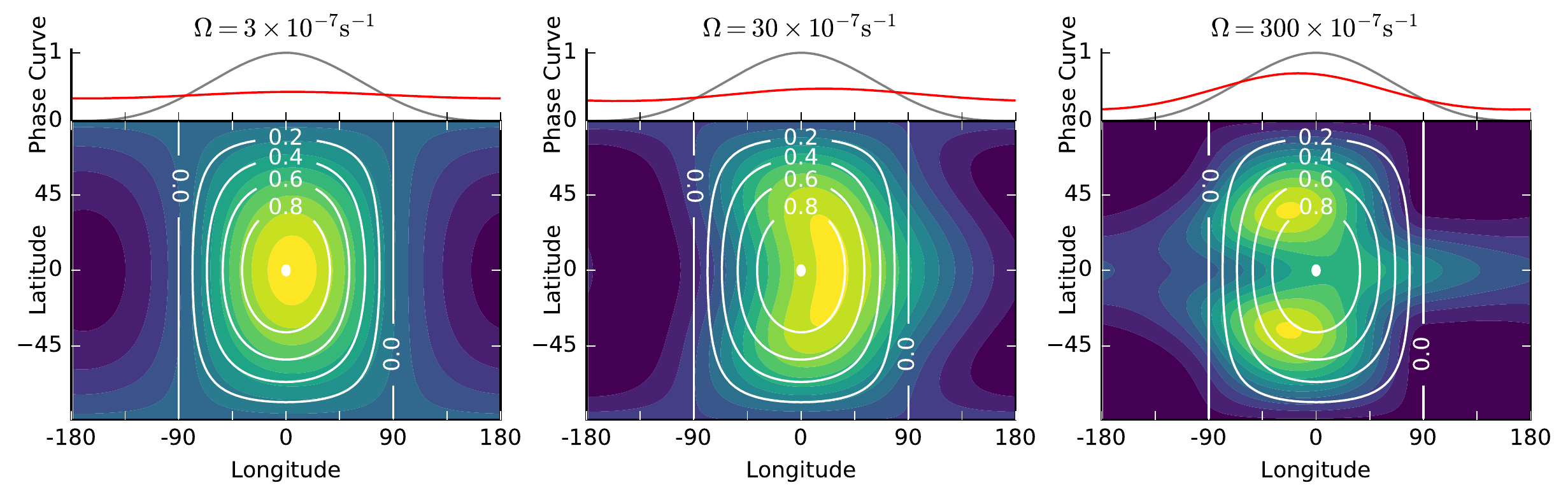}
	\caption{Steady-state solutions for tidally-locked ($\alpha=0$) forcing at increasing planetary rotation rate, with corresponding integrated phase curves. This subset corresponds to a transpose of the central column of Figure~\ref{fig:omega_spin_space}.
	Lower panel surface plots show, as in Figure~\ref{fig:omega_spin_space}, $\heq$ in white contours and the steady-state $h$ in the ($\xi$, $\phi$) reference frame.
	Corresponding normalised phase curves are shown in the top panel for the steady-state height field $h$ (red line) and the equilibrium field $\heq$ (grey line).  $gH=\SI{1000}{m^{2}s^{-2}}$, $\tau = \SI{5}{days}$.}
	\label{fig:tl_vary_omega}
\end{figure}
Figure~\ref{fig:tl_vary_omega} shows the height field $h$ and the corresponding normalised phase curve of three tidally-locked ($\alpha = 0$) runs with increasing $\Omega$.
In the tidally-locked case forcing is stationary in both $\lambda$ and $\xi$ and the effect of increasing rotation rate is isolated.

In the slowest rotating case ($\Omega = \SI{3e-7}{\per\second}$) planetary scale divergent flow, damped over timescale $\tau$, balances the height gradient.
Day-night temperature differences are minimal; this response is similar to the weak temperature gradient (WTG) solution of \cite{Bretherton:2003eh} over a planetary scale and the direct day-night circulation observed by \cite{Kato:1994wb}.
The amplitude of the phase curve is small, with a prograde offset induced in the tidally-locked configuration by a small amplitude trapped-Kelvin wave.

As $\Omega$ increases the Rossby deformation radius, $L_d$, decreases and flow becomes dominated by rotational-effects.
In the integrated phase curve, the change in character of the solution results in shift in the maximum from an easterly to westerly offset from the substellar point -- the contribution from trapped Rossby waves in the subtropics to the west of the substellar point becomes the major feature of the geopotential field.
With faster rotation the flow becomes more geostrophic; the amplitude of the phase curve increases as a larger Coriolis force balances a larger temperature gradient.
\begin{figure}[tb]
	\centering
	\includegraphics[width=0.45\textwidth]{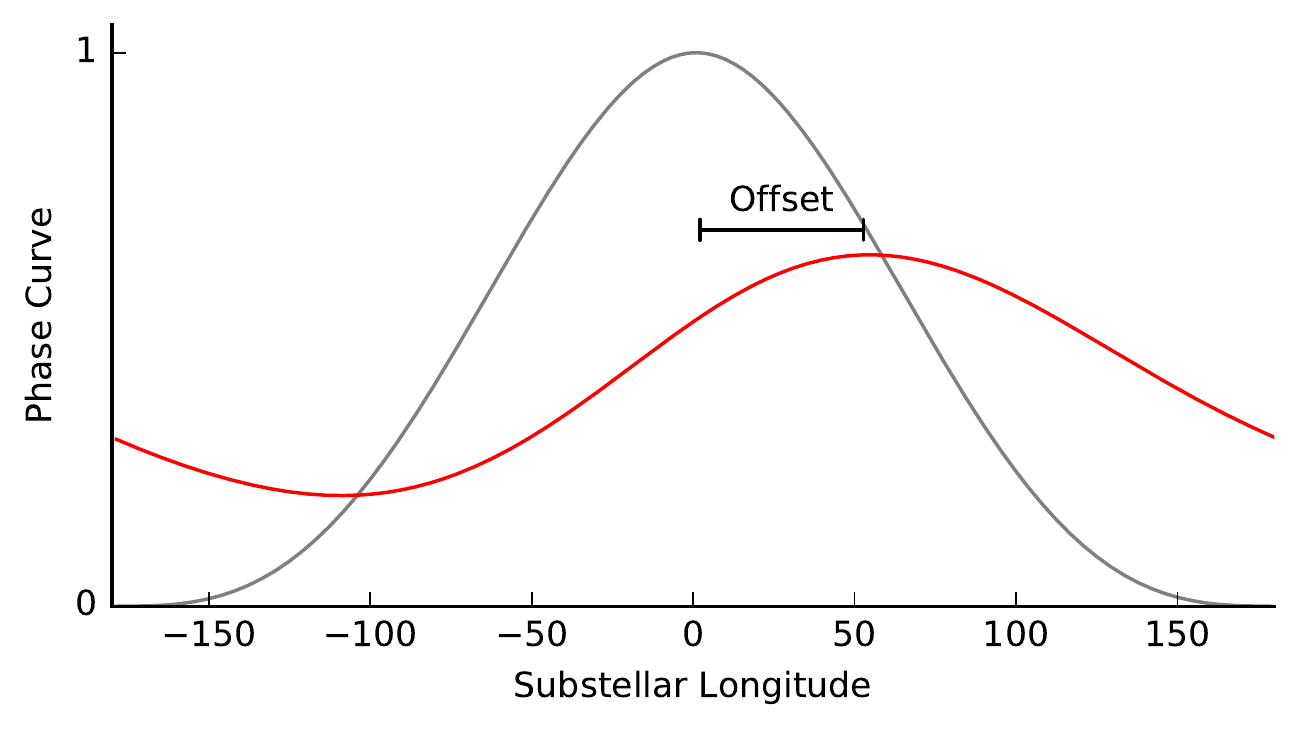}
	\caption{Example normalised phase curve of the equilibrium profile $\heq$ (gray line) and a steady-state height field (red line).
	The \emph{phase curve offset} is the longitudinal distance from the centre of the forcing to the maximum of the integrated phase curve, marked above.
	}
	\label{fig:phase_curve_labelled}
\end{figure}

\begin{figure}[tb]
	\centering
	\includegraphics[width=\textwidth]{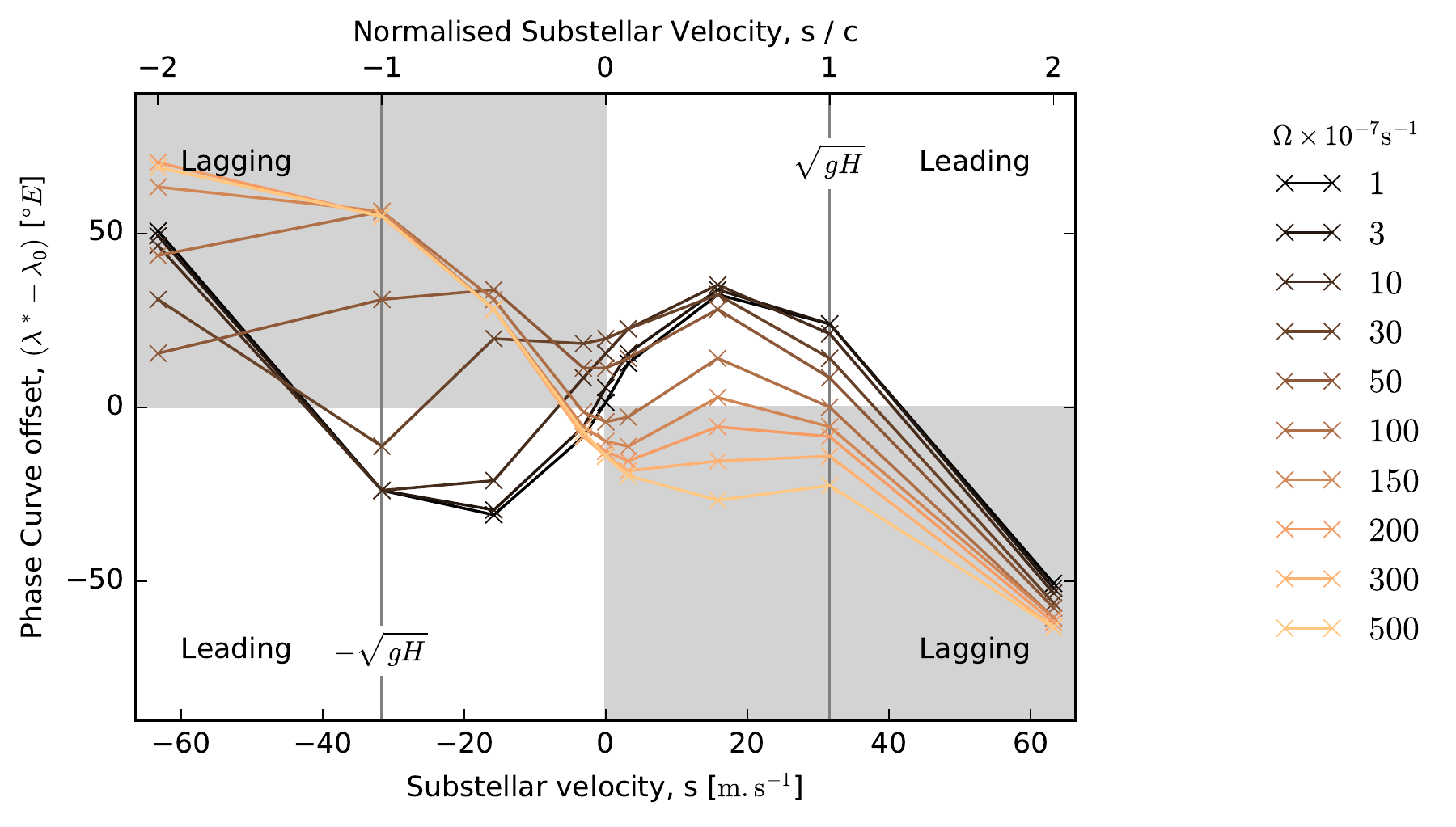}
	\caption{Phase curve offset for varying substellar velocity.  Increasing brightness lines show the magnitude of offset as rotation rate $\Omega$ is increased from \SI{1e-7}{s^{-1}} to \SI{5e-4}{s^{-1}}.
	The phase curve offset is measured from the substellar point at $\xi = 0$.  The point at which the substellar point is moving at Kelvin wavespeed, $c = \sqrt{gH}$ ($\alpha = 1$), in either the prograde and retrograde direction, is marked with a vertical line.  Shaded regions mark where the hotspot lags behind the substellar point. i.e. If the substellar point is moving in a prograde direction, $\lambda_0$~velocity~$> 0$, the hotspot offset is negative.
	$gH = \SI{1000}{m^2 s^{-2}}$, $\tau = \SI{5}{days}.$}
	\label{fig:pc_offset_vs_vel}
\end{figure}

For the parameter space varying $\Omega$ and $\alpha$ we consider the offset of the integrated phase curve from the substellar point.
The phase curve offset is defined as the longitudinal distance of the maximum of the phase curve from the centre of the forcing function (Figure~\ref{fig:phase_curve_labelled}).
Figure~\ref{fig:pc_offset_vs_vel} shows the magnitude of the offset for increasing substellar velocity, $s$, at increasing rotation rates.

In the slowly rotating limit, offsets converge to a curved profile (See $\Omega= \num{1}, \num{3}, \SI{10e-7}{s^{-1}}$ lines in Figure~\ref{fig:pc_offset_vs_vel}) with the hotspot preceding the substellar point with a peak at $|\alpha|  \simeq  \pm 1/2$ before lagging once $|\alpha| > 3/2$.
In the fast rotation case, the hotspot lags behind the substellar point in all but slowest retrograde substellar motion.

For prograde substellar motion, $0 \le \alpha \le 1$, there is a smooth transition from a leading to lagging hotspot as planetary rotation rate increases.
The 5th and 6th columns of Figure~\ref{fig:omega_spin_space} show corresponding global geopotential profiles -- as the $\Ro$ number decreases down the column planetary waves become more apparent in the steady-state and the latitudinally integrated peak response moves from being prograde (leading) to retrograde (lagging) the substellar point.

\begin{figure}[tb]
	\centering
	\includegraphics[width=\textwidth]{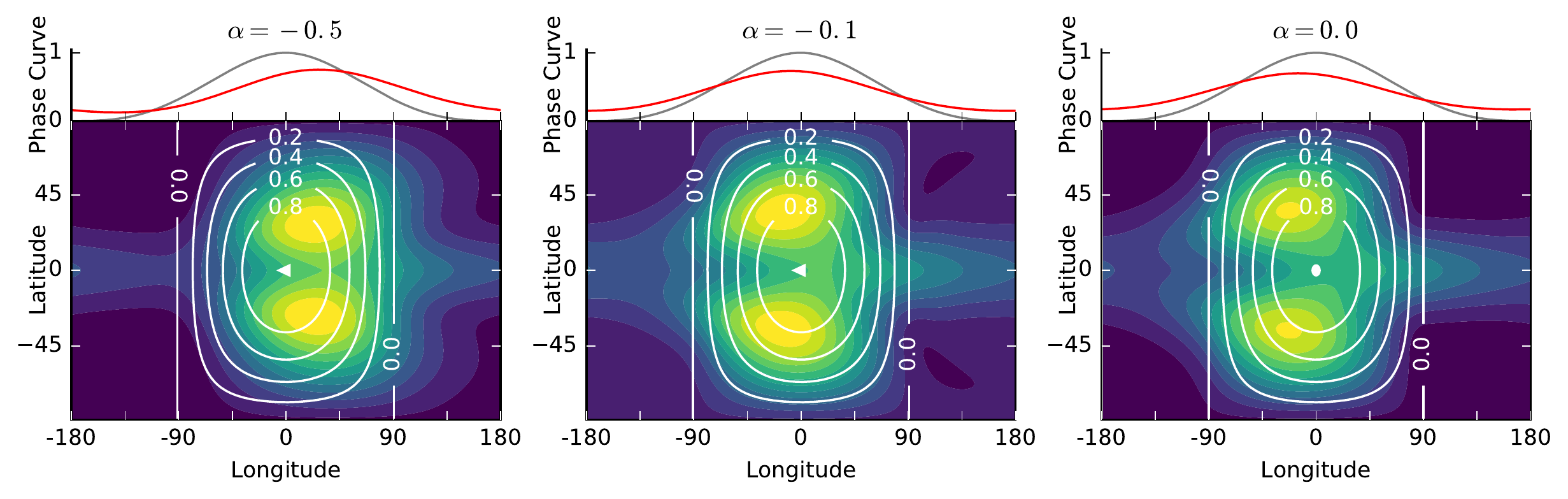}
	\caption{Steady-state geopotential and integrated phase curves of fast rotation and slow westward forcing propagation.  The initially westward offset of the Rossby gyres when $|\alpha| < 0.5$ becomes eastward once the forcing speed becomes faster than the planetary wave speed. $\Omega = \SI{300e-7}{\per\second}$, $\sqrt{gH} = \SI{10}{m s^{-1}}$, $\tau = \SI{5}{days}$.}
	\label{fig:slow_rossby}
\end{figure}

Compare this to retrograde substellar motion, $-1 \le \alpha < 0$, where the offsets in fast and slow rotating experiments separate into two distinct solutions.
The prograde propagating Kelvin waves and gravity waves move eastward with wavespeed $c\simeq\sqrt{gH}$ while the wavenumber-1 westward propagating planetary wave travels at $1/3$ the velocity of the Kelvin wave \citep{Gill:1980bma}.
On slowly rotating planets ($\Ro \gtrsim 1$) the deformation radius is larger than the scale of forcing and thus planetary waves are not observed -- the Coriolis gradient is insufficient to support them.
However on fast rotating planets where planetary waves do occur, the difference in velocity of the two wave modes results in phase curve offsets that are distinct for a substellar point that propagates to the east or west.
For fast rotation with a stationary forcing the Rossby gyres lie west of the centre of the substellar point and the integrated phase curve has a westerly offset.
When the substellar point is moving slowly retrograde ($-1/3 < \alpha < 0$) the westerly offset is maintained, but as the speed of the substellar point increases beoyond this the planetary waves lag behind the forcing.
Figure~\ref{fig:slow_rossby} shows the steady-state geopotential and phase curves for $\alpha = 0, -0.1, -0.5$ in a fast rotating system, the Rossby gyres clearly shifting from leading to lagging the substellar point as retrograde substellar motion increases.

When the substellar point is moving prograde the impact of the planetary waves on the phase curve is less pronounced.
As the substellar prograde velocity transitions to being faster than the Rossby wave group speed the gyres become progressively more longitudinally smoothed out across the domain, damped by the radiative cooling process (see lower right of Figure~\ref{fig:omega_spin_space}).
At high eastward substellar velocity, $\alpha > 1/3$, the hotspot returns to being on the equator; the adjustment to balancing the moving forcing dominating over the rotational dynamics.  These results may be compared with those of the slowly rotating experiments where the deformation radius is larger than the scale of forcing and planetary waves are not observed.

\subsection{The effect of Rayleigh and Newtonian damping} 
\label{sub:magnitude_of_offset}

\begin{figure}[tb]
	\centering
	\includegraphics[width=0.8\textwidth]{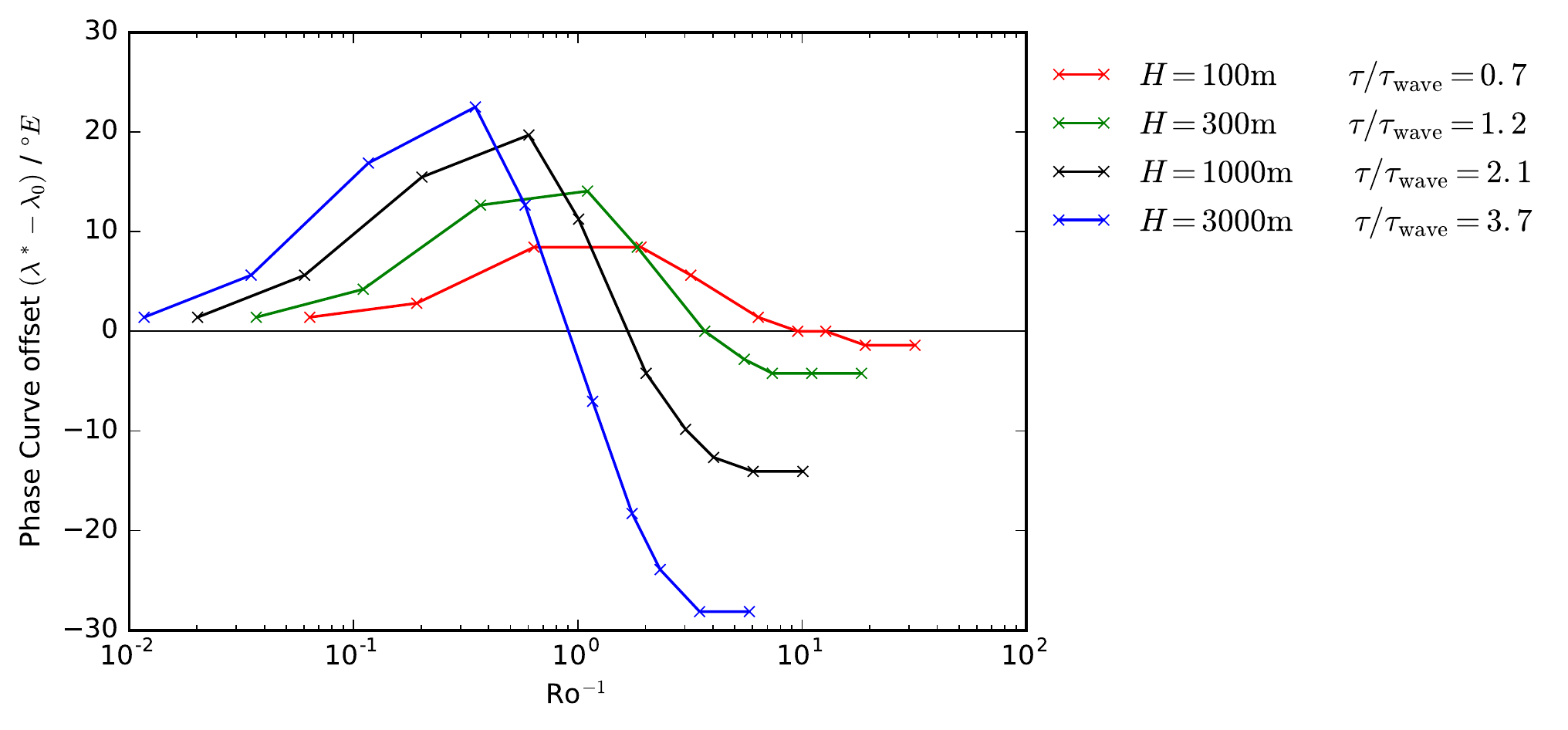}
	\caption{Tidally-locked hotspot offset with varying fluid depth and constant $\tau = \SI{5}{days}$, as a function of inverse planetary Rossby number.  The curves show results with different fluid depths, $H$,  with a constant frictional timescale $\tau=\SI{5}{days}$, such that the ratio $\tau/\twave$ increases with increasing fluid depth. As $\tau/\twave$ increases, the magnitude of the tidally-locked hotspot offset increases and the transition from eastward to westward offset occurs at a slower rotation rate; that is, at a smaller inverse planetary Rossby number.
	Compare to Figure~\ref{fig:offset_by_h_matched_tau} where the ratio $\tau/\twave$ is kept constant at a value of 2.1.}
	\label{fig:offset_by_h_5d_tau}
\end{figure}

\begin{figure}[tb]
	\centering
	\includegraphics[width=0.8\textwidth]{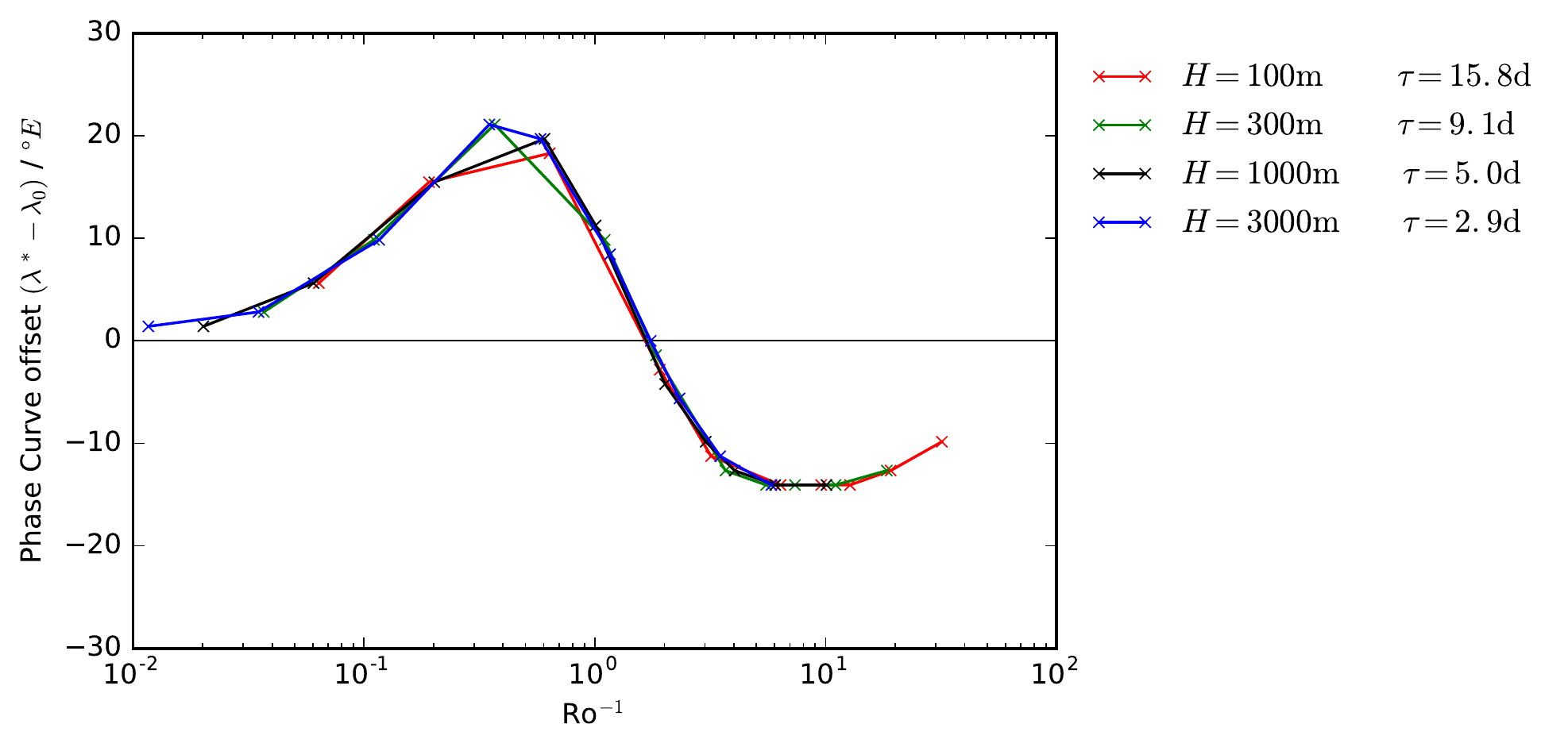}
	\caption{Tidally-locked hotspot offset with varying fluid depth and constant $\tau/\twave = 2.1$ as a function of inverse planetary Rossby number. The various curves show results with different
	fluid depths, $H$, but with drag timescale ratio kept constant, $\tau/\twave = 2.1$ (using $H = \SI{1000}{m}$, $\tau = \SI{5}{days}$ as a reference, black line).
	The hotspot offset observed is consistent across all experiments with the same inverse planetary Rossby number.}
	\label{fig:offset_by_h_matched_tau}
\end{figure}

In our dimensional results presented above we have made a choice of Earth-like values $\tau = \SI{5}{days}$, $gH = \SI{1000}{m^2 s^{-2}}$, $a = \SI{6317}{km}$.
In studying exoplanets, we wish to expand our investigation beyond the parameter regime of Earth, and so we now consider varying both $\tau$ and $H$.
Here we will show that it is the non-dimensional frictional timescale $\tau/\twave$ that determines the influence of frictional forces on the offset of the phase curve.  Maintaining a constant $\tau$ and reducing fluid depth $H$, hence slowing the wavespeed in the system and decreasing $\tau/\twave$, means stronger damping relative to the wave motion and thus a reduction in the magnitude of the hotspot offset (Figure~\ref{fig:offset_by_h_5d_tau}).
If the frictional/radiative damping is scaled in proportion, keeping $\tau/\twave$ constant and varying fluid height $H$ an equivalent response is observed for the same Rossby number (Figure~\ref{fig:offset_by_h_matched_tau}).

Here we show for comparison the effect on varying $H$ on the offset of a tidally-locked system, however the qualitative results extend to the full range of $\alpha$ values shown in Figure~\ref{fig:pc_offset_vs_vel} -- as $\tau/\twave$ gets smaller the phase curve offset becomes smaller -- the impact of shorter frictional timescales is stronger relaxation towards the forcing and stronger damping of advective fluid velocity, resulting in a steady-state height field that is phase-aligned with the forcing.  The constancy of results for constant $\Ro$, and $\tau/\twave$ show that it is the scale of the frictional forces relative to the wavespeed of the atmosphere that influence the observed phase curve, not the absolute value.

\subsection{Substellar motion induced offset in the slow rotating limit} 
\label{sub:slowly_rotating_planets}
To understand the offset in the slow limit we can consider an even simpler model, one that captures the fundamental behaviour of the hotspot preceding the motion of the substellar point.

\begin{figure}[tb]
	\centering
	\includegraphics[width=0.8\textwidth]{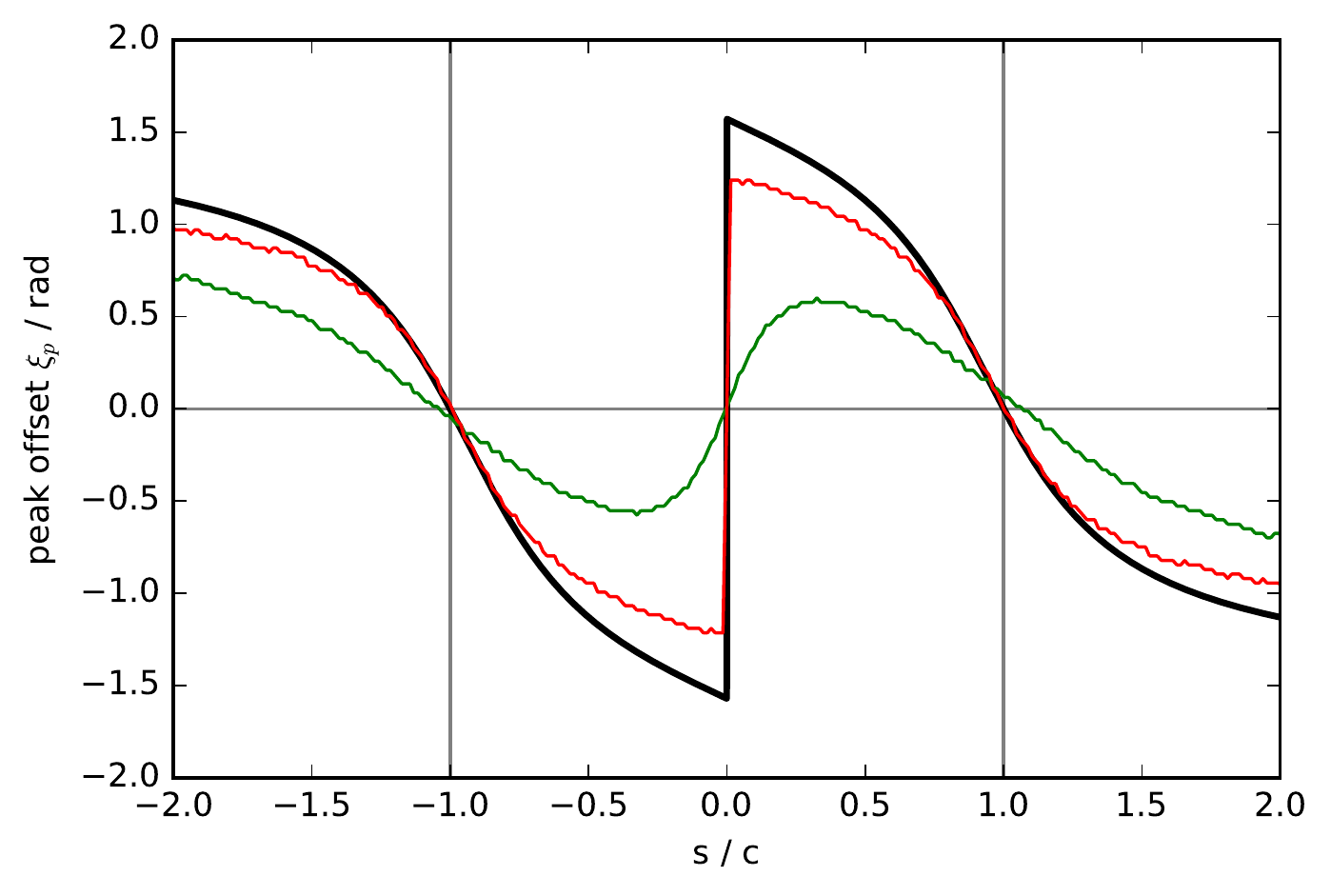}
	\caption{Analytic and numerical phase curve offset in a one dimensional shallow water model.  The thick black line is the analytic offset given by \eqref{eqn:h_anal_offset}.  The red line shows the numerical integration of equations \eqref{eqn:one-dim-u}--\eqref{eqn:one-dim-h} with $\heq = \max(\cos\xi, 0)$, reducing the the amplitude of offset compared to the full cosine forcing.
	When non-linear terms and Rayleigh drag are additionally included (green line), the velocity of zero offset becomes faster than linear wavespeed and the transition through the origin is smoothed producing in a response that approximates the results of the slowly-rotating non-linear two-dimensional model in Figure~\ref{fig:pc_offset_vs_vel}.}
	\label{fig:one-dim_numerical}
\end{figure}

Consider the one-dimensional, non-rotating shallow water equations linearised about a basic state $u = 0 + u'$, $h = H + h'$ and with a simple forcing analogous to (\ref{eqn:model_heq}):
\begin{align}
	\dd{u'}{t} + g \dd{h'}{x} &= 0, \label{eqn:one-dim-u} \\
	\dd{h'}{t} + H \dd{u'}{x} &= \frac{\heq - h'}{\tau}, \label{eqn:one-dim-h}  \\
	\heq &= \Delta h \cos(x - st),
\end{align}
where $s$ is the speed of propagation of the forcing and $\Delta h$ is the scale of a sinusoidal forcing.
We do not include a Rayleigh frictional term in the velocity evolution equation as here we intend to isolate the effect of the moving forcing from the other balancing forces present in the system.
This is equivalent to taking a view along the equator of a non-rotating planet modelled by equations \eqref{eqn:model_u} -- \eqref{eqn:xi}.
Since $\vect f = 0$ and the equations are symmetric in latitude, at the equator $\partial /\partial {y} \equiv 0$ and we can reduce the problem to the one-dimensional case.
For simplicity of exposition, here we use a full cosine wave rather than the slightly more complex $\heq = \max(\cos(x - st), 0)$ that would be a true one-dimensional equivalent to \eqref{eqn:model_heq}.
Numerical modelling of this half-cosine wave one-dimensional forcing provided offset results almost identical to the analytic solution -- see Figure~\ref{fig:one-dim_numerical} for comparison.

As for the rotating two-dimensional system, we again introduce a coordinate system that moves with the forcing.
Defining $\xi = x - st$ we can consider the steady-state solutions of \eqref{eqn:one-dim-u} -- \eqref{eqn:one-dim-h}.
Using the transform identities
\begin{equation}
	\dd{}{x} \to \diff{}{\xi} \quad \quad \dd{}{t} \to -s\diff{}{\xi}, \label{eqn:change_coords}
\end{equation}
and upon dropping primes \eqref{eqn:one-dim-u}--\eqref{eqn:one-dim-h} become ordinary differential equations
\begin{align}
	-s\diff{u}{\xi}& + g \diff{h}{\xi} = 0, \\
	-s\diff{h}{\xi}& + H \diff{u}{\xi} - \frac{\Delta h\cos\xi}{\tau} + \frac{h}{\tau} = 0,
\end{align}
that can be combined to give a single expression for $h$
\begin{equation}
	\left(\frac{gH}{s} - s\right) \diff{h}{\xi} + \frac{h}{\tau} - \frac{\Delta h\cos\xi}{\tau} = 0. \label{eqn:h_one-dim-1}
\end{equation}
Using the same gravity wavespeed established in the two-dimensional case, $c = \sqrt{gH}$, we write (\ref{eqn:h_one-dim-1}) as
\begin{equation}
	\frac{(c^2 - s^2)}{s} \diff{h}{\xi} + \frac{h}{\tau} - \frac{\Delta h\cos\xi}{\tau} = 0, \label{eqn:h_one-dim-2}
\end{equation}
which can be solved using an integration factor $I = e^{a \xi}$, with $a = s/(\tau(c^2 - s^2))$ to obtain an analytic solution
\begin{equation}
	h(\xi) = \frac{\Delta h a(a \cos \xi + \sin \xi)}{a^2 + 1}. \label{eqn:h_anal}
\end{equation}
The one-dimensional peak offset $\xi_p$ can be found from the inflection point, where $\diff{h}{\xi}(\xi_p) = 0$.
Taking the derivative of (\ref{eqn:h_anal})
\begin{equation}
	\diff{h}{\xi} = \frac{\Delta h a(\cos \xi - a \sin \xi)}{a^2 + 1} \label{eqn:h_anal_diff}
\end{equation}
and equating to zero we can obtain an analytic solution for the peak offset in the one-dimensional case
\begin{align}
	a \sin \xi_p &= \cos \xi_p, \\
	\implies \xi_p &= \arctan\left(\frac{(c^2 - s^2)\tau}{s}\right). \label{eqn:h_anal_offset}
\end{align}

Figure~\ref{fig:one-dim_numerical} shows the peak offset (\ref{eqn:h_anal_offset}) as a function of forcing velocity $s$, as well as numerical solutions of the one-dimensional system using a non-smooth forcing and with non-dimensional terms and a Rayleigh drag term included on the zonal velocity.
When the forcing is slow moving ($|s| \ll c$) the height field peaks ahead of the forcing by a factor of $\pi/2$ (since $\arctan(x) \to \pm\pi/2$ as $x \to \pm \infty$), smoothly transitioning to a lagging phase as forcing speed exceeds wavespeed.
The lagging offset observed from very fast moving substellar point will also tend to a limit of $\mp\pi/2$ when $\pm s \gg c$.
At $s = 0$ there appears to be a sharp discontinuity in the location of the peak as it transitions from a westward to eastward propagating forcing.
This discontinuity is not observed;
in the linear model the amplitude of the steady-state fluid height goes to zero in the tidally-locked case, as is clear from (\ref{eqn:h_anal}) where coefficient $a \to 0$ as $s \to 0$.
In our more complex model that retains the non-linear terms and includes a Rayleigh drag term, the transition through the tidally-locked state is smooth, as shown in Figure~\ref{fig:one-dim_numerical}.
This can be compared to the slowly-rotating two-dimensional offsets in Figure~\ref{fig:pc_offset_vs_vel}.

%% file: discussion.tex
\pagebreak[4]
\section{Discussion and Interpretion} 
\label{sec:discussion}
\subsection{Dynamical balances}
A gradient in the height field can only be maintained by balancing forces, and by considering these balances we can intepret the results.
In the linearised form of (\ref{eqn:u}), and at steady-state in the coordinate frame ($\phi$, $\xi$) (using \eqref{eqn:xi} to perform a change of coordinates), we see that a gradient in the height field could be maintained by three sources, being the left-hand side terms the momentum equation
\begin{equation}
	-\frac{s}{a} \dd{\vect u}{\xi} + \vect f \times \vect u + \frac{\vect u}{\tau} = -\nabla gh .
    \label{mom:bal}
\end{equation}

Depending on the parameter regime, defined by planetary Rossby number $\Ro = \sqrt{gH}/\Omega a$, substellar velocity $s$, and the influence of frictional forces $\tau/\twave$, we consider the case where each of the terms on the left is the dominant balancing force:
\begin{enumerate}
	\item Geostrophic balance: the gradient in height is maintained by the Coriolis force
	\begin{equation}
		\vect f \times \vect u = -\nabla gh.
	\end{equation}
	In fast rotating systems, $\Ro \ll 1$, this balance will dominate the flow. Figure~\ref{fig:offset_by_h_matched_tau} shows the onset of geostrophic balance in the tidally-locked case -- as $Ro^{-1} > 1$ the influence of Rossby gyres moves the offset westwards of the forcing.
	\item Frictional balance: the gradient is balanced by drag forces
	\begin{equation}
		\frac{\vect u}{\tau} = -\nabla gh.
	\end{equation}
	As $\tau/\twave$ becomes smaller, this becomes the dominant balance.
	Strong damping of the velocity field prevents redistribution of momentum around the domain, and the steady-state height field becomes phase-aligned with the forcing, reducing the offset created by both Kelvin and Rossby waves (Figure~\ref{fig:offset_by_h_5d_tau}).
	\item Dynamical balance: the gradient is balanced by motion in the forcing
	\begin{equation}
		-\frac{s}{a} \dd{\vect u}{\xi} = -\nabla gh.
	\end{equation}
	Most evident in slowly rotating planets ($\Ro \gtrsim 1$) the dynamical balance introduced by a moving forcing can dominate the phase offset when $|s| < c$.
	In this regime the hotspot can preceed the motion of the substellar point, the height field will be up to $\pi/2$ out of phase with the forcing to establish a steady-state solution (Figure~\ref{fig:one-dim_numerical}).
\end{enumerate}

\begin{figure}
\gridline{
	\fig{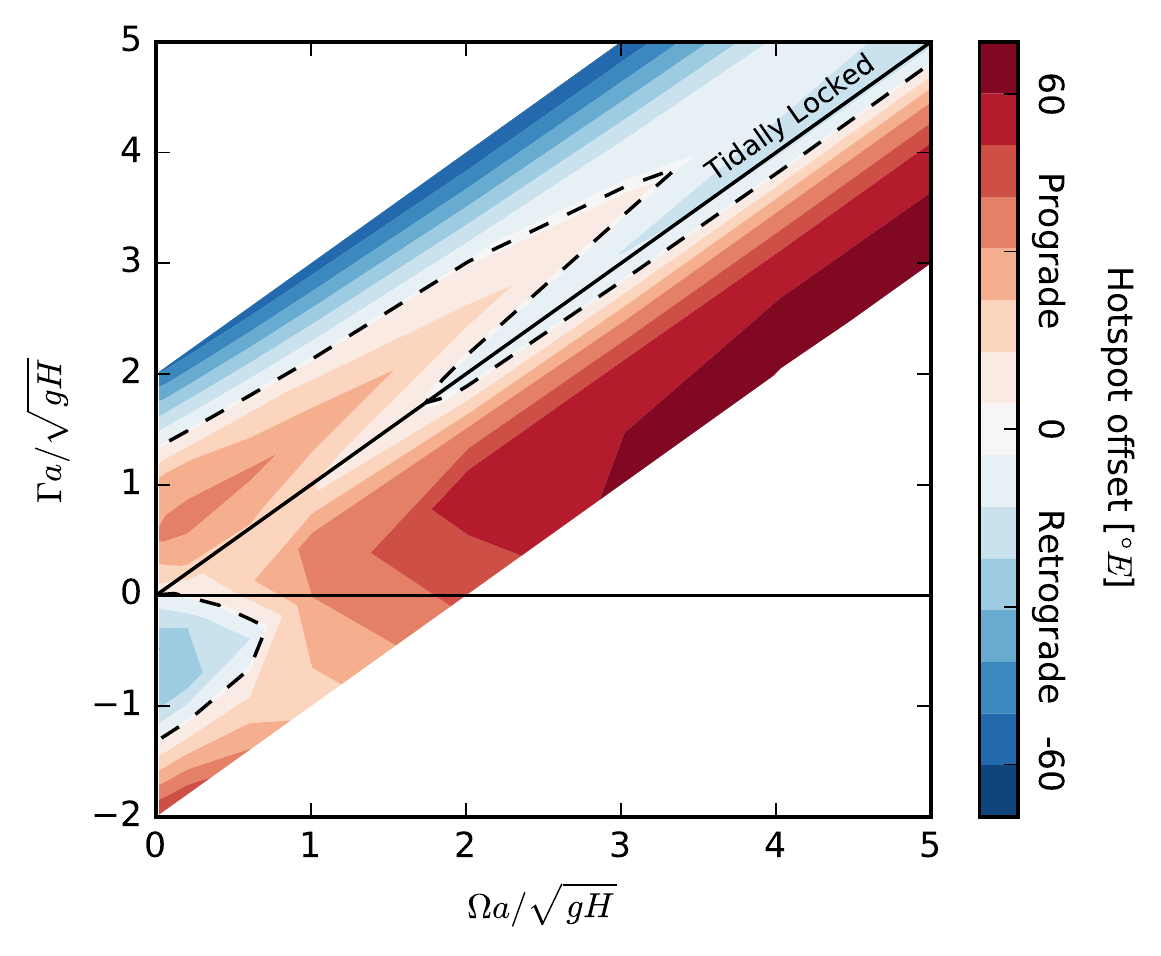}{0.45\textwidth}{(a) Hotspot offset: prograde or retrograde}
	\fig{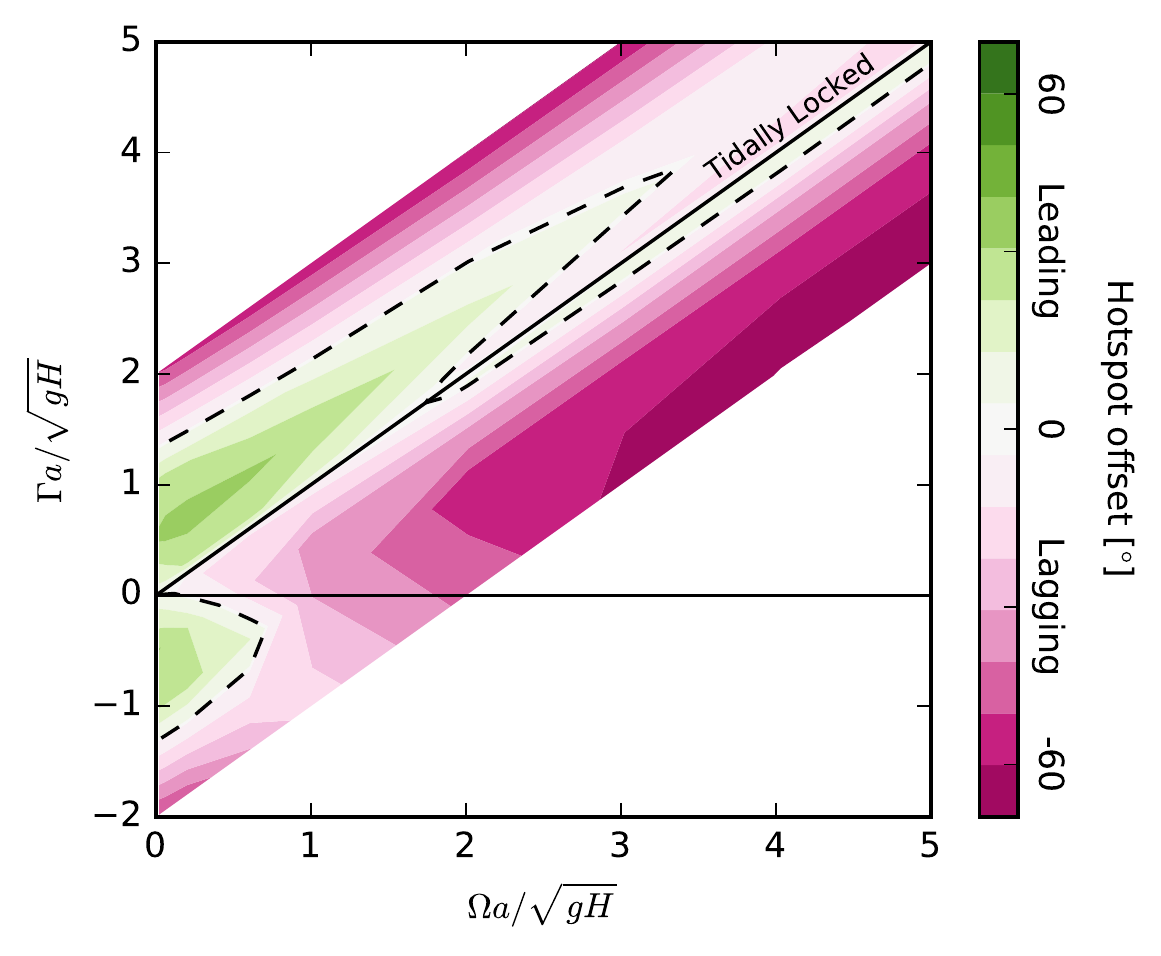}{0.45\textwidth}{(b) Hotspot offset: leading or lagging }}
\gridline{
	\fig{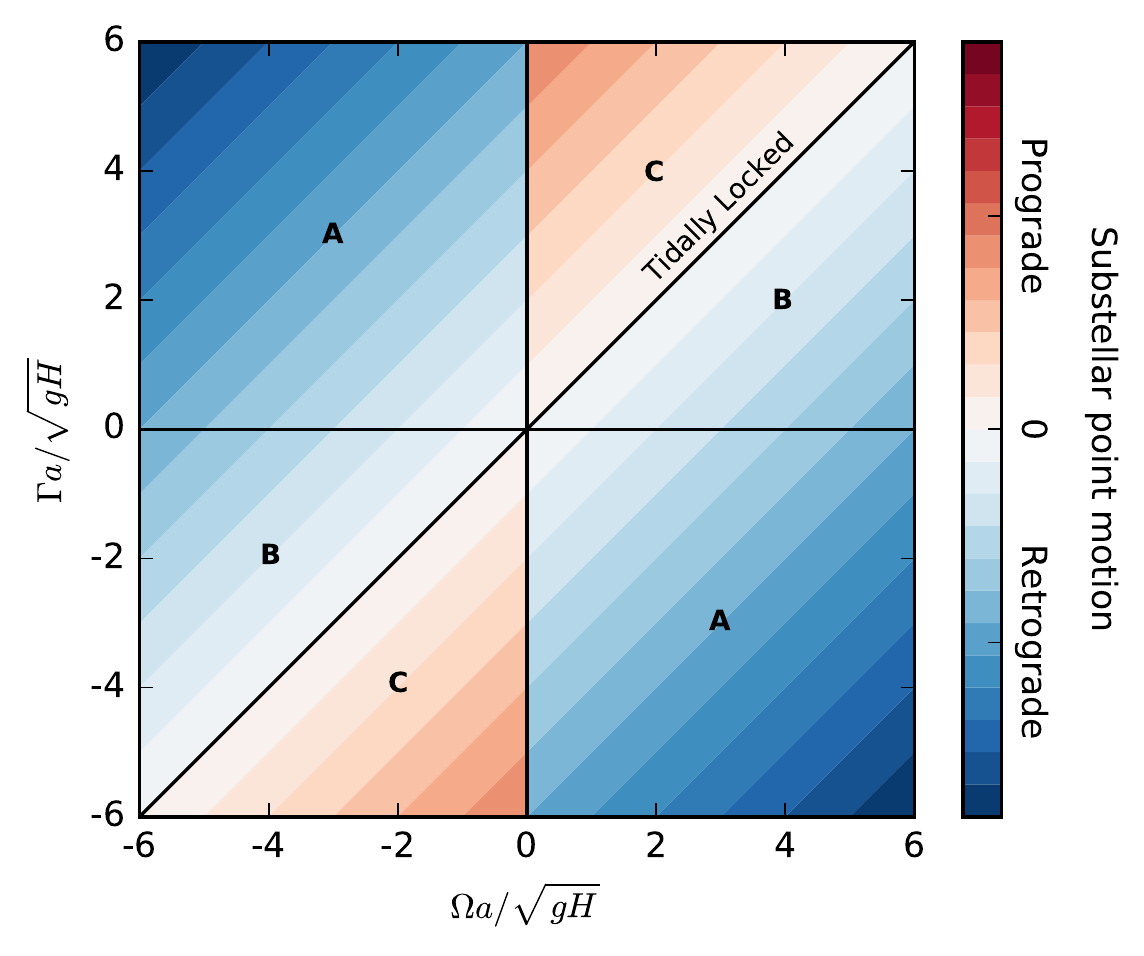}{0.45\textwidth}{(c) Substellar point velocity}
    \fig{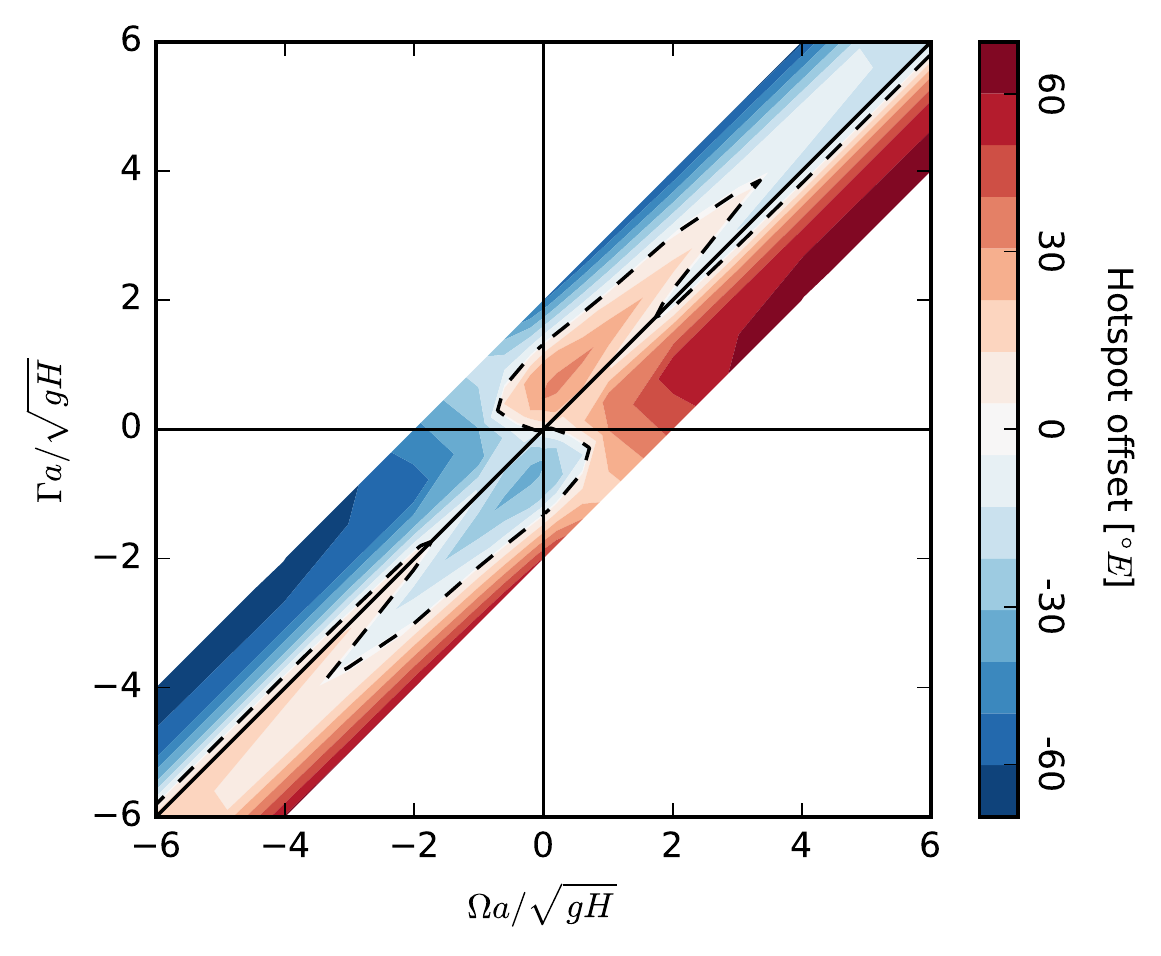}{0.45\textwidth}{(d) Hotspot offset}}
\caption{Regime diagrams of substellar point velocity and hotspot offset as a function of orbital rate, $\Gamma$, and rotation rate, $\Omega$.
A planet is tidally-locked when $\Omega = \Gamma$ -- marked by a black line along the diagonal.
The upper panel shows the hotspot offset given by the shallow water model in either the refrence frame of (a) prograde/retrograde offset, relative to the rotation vector, or (b) leading/lagging offset, relative to the motion of the substellar point.
In the lower panel, (c) plots substellar point velocity from (\ref{eqn:substellar_vel}).
(d) shows the hotspot location, relative to the substellar point, for the complete $(\Omega, \Gamma$) space.
The zero contour -- when the hotspot is at the substellar point, is shown with dashed-black line on figures (a), (b), (d).
Empty regions far from the diagonal are outside the range of substellar velocities tested.
}
\label{fig:substellar_motion}
\end{figure}

\subsection{Celestial Mechanics}
In the construction of the model, we divorced the motion of the substellar point from the celestial mechanics that induce the diurnal cycle, now here we would like to bring the results back into the context of an orbiting exoplanet.
Whether the substellar point moves in a prograde or retrograde direction relative to planetary rotation depends on both the direction and magnitude of the rotation rate, $\Omega$, and orbital rate, $\Gamma$, as given in equation~(\ref{eqn:substellar_vel}).

Figures~\ref{fig:substellar_motion} (a) and (b) are transformations of Figure~\ref{fig:pc_offset_vs_vel}, showing the hotspot offset predicted by the model for an observed exoplanet where the orbital rate is derived from (\ref{eqn:substellar_vel}),
\begin{equation}
	\Gamma = \frac{\alpha c}{a} - \Omega.
\end{equation}

Figure~\ref{fig:substellar_motion} (c) maps the substellar point velocity over the ($\Omega$, $\Gamma$) parameter space; speed increases away from the tidally-locked diagonal where $\Omega = \Gamma$.
When the planet rotates in the opposite direction to its orbit the substellar motion is always retrograde (regions A);
when the planet and orbit move in the same direction there are two possible regimes.
If the planet is rotating faster than its orbital rate (regions B) then here too the substellar point moves retrograde to the planetary rotation.
This is the regime in which Earth lies, $\Omega_\Earth > \Gamma_\Earth > 0$ such that the sun appears to rise in the east and set in the west.
Only when $|\Gamma| > |\Omega|$ and $\Omega\Gamma > 0$ (regions C) will the substellar point move in the same direction as the rotation of the planet.

In this study by varying $\alpha$ and with $\Omega > 0$ we have considered only the right-half plane of Figure~\ref{fig:substellar_motion}(c).
Due to the longitudinal symmetry of equations (\ref{eqn:model_u})--(\ref{eqn:xi}) the same dynamics will hold true for planets rotating in the other direction,  appropriately reflected in the longitudinal direction to account for the reversal of direction of the Coriolis vector and we can use our results to fill this space also, as shown in Figure~\ref{fig:substellar_motion}(d).

\begin{figure}[tb]
	\centering
	\includegraphics[width=0.55\textwidth]{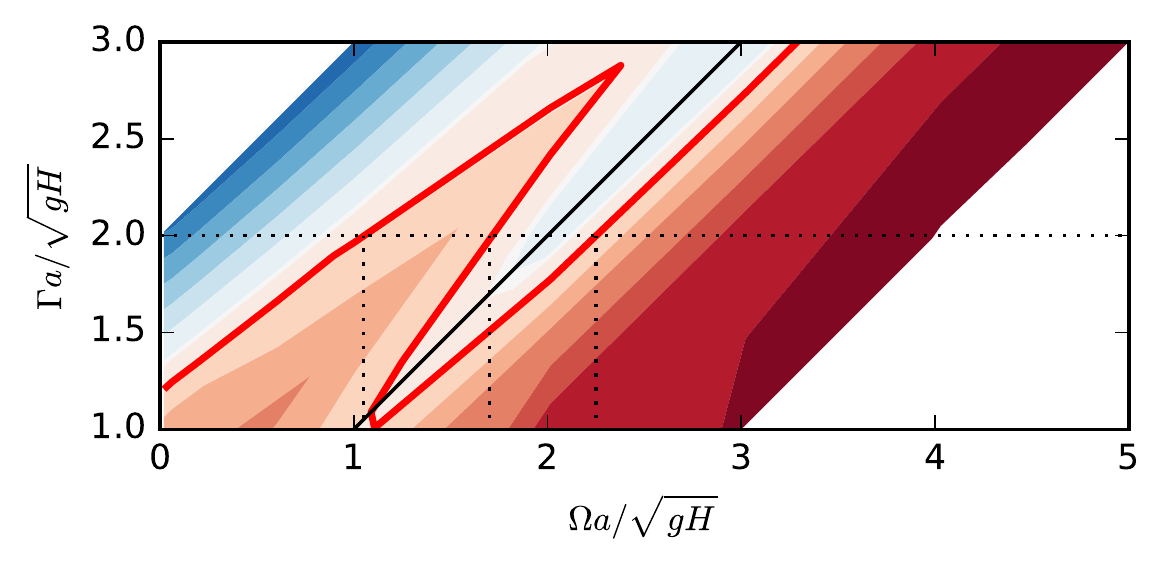}
	\caption{Blow up of Figure~\ref{fig:substellar_motion} (a) around $\Gamma a / \sqrt{gH} = 2$.
	Given an observed orbital rate $\Gamma a / \sqrt{gH} = 2$ and phase curve offset of 10ºE (red contour), the shallow water model provides three possible rotation rates (vertical black dotted lines).}
	\label{fig:xsection}
\end{figure}

Using transit detections and the spin direction of the star it been shown that we can constrain the direction and period of the orbit of a planet \citep{Queloz:2000ui} and so calculate $\Gamma$.
The results presented above suggest that fully constraining the rotation rate from a known orbital rate and the phase curve may not be possible even when considering only a shallow water model of the atmosphere, but some useful constraints nevertheless emerge. For example, as substellar velocity increases in either direction over the planet, the hotspot tends to a lagging limit. For exoplanets that exhibit very large hotspot offsets and lie outside an orbital radius of certain tidal-locking, it may be possible to constrain the direction of rotation from the phase curve offset.
But for planets that are tidally-locked or near-locked, as described for Figure~\ref{fig:pc_offset_vs_vel} above, the hotspot can potentially lie east or west of the substellar point.
Thus, zooming in around $\Gamma a / \sqrt{gH} = 2$ on Figure~\ref{fig:substellar_motion} (a),
an observed offset of 10ºE (shown as a red contour on the figure) could be attributed to three different rotation rates (Figure~\ref{fig:xsection}).
However, at this orbital and rotation rate the model does predict that a tidally-locked planet should have a small westward phase curve offset, so although such an observation may not provide a tight constraint on rotation rate it can tell us that the planet is not tidally-locked.

The magnitude of phase curve offset in the shallow water model is highly dependent on the damping timescale $\tau/\twave$ and the atmospheric wave speed $\sqrt{gH}$. These parameters have analogues in more complex three-dimensional treatments of a planetary atmosphere -- frictional diffusivity, thermal radiation to space, the Brünt-Vaisalla frequency. These properties of the fluid and will vary greatly depending on the chemical composition and scale height of the atmosphere and are not easily determined from remote observations.  Spectroscopic measurement of atmospheric mass and composition could lead to tighter constraint on the radiative timescale of an exoplanet atmosphere and potentially too the scale height, from which a ratio could be established and inference made.


\section{Conclusions} 
\label{sec:conclusions}
We have used a shallow water model to demonstrate that the large-scale dynamics of exoplanetary atmospheres are sensitive to both the rotation rate of the planet and the length of its diurnal cycle.
When considering this from the view of an observer, these changes are manifested in the peak of the thermal phase curve being offset from the point of maximal stellar insolation.
Even in the simplest case of a tidally-locked planet, the peak of the observed phase curve can be either east or west of the substellar point, depending on the rotation rate and scale height of the atmosphere, as has been previously demonstrated. When the substellar point is moving across the atmosphere of a planet, the effect on the atmospheric dynamics additionally depends on the speed of the motion of the forcing at the equator, $s$, compared with the internal wavespeed of the atmosphere, $c$. When $|s| > c$ the hottest point of the atmosphere lags behind the substellar point; in the reference frame of a fixed point in the atmosphere of the planet this means it gets hottest after midday, once the stellar zenith has passed.

For planets with a long diurnal cycle where the substellar point moves slowly over the surface, $|s| < c$, the speed of substellar motion has a dominant impact on the phase curve, the peak can be ahead of the stellar zenith -- the hottest point of the day would be in the morning, before the point of maximal stellar irradiation. This is potentially a pertinent area of parameter space for terrestrial exoplanets and Venus falls into this regime.

Due to the degeneracy of multiple balancing forces at play in the maintenance of a global temperature gradient, the shallow-water model does not provide a tight constraint on the rotation rate of a planet from the phase curve offset, and the model also omits a number of effects associate with stratification that will likely modulate the dynamics and thermal profile. However, Kelvin wave and Rossby wave dynamics are very robust phenomena that carry over into a fully stratified atmosphere, and understanding the effect of asynchronous rotation on the large-scale dynamics in a simple model such as this will in any case provide insight into the more complex atmospheric circulations observed in three-dimensional GCMs.

Perhaps the most significant limitation of the shallow water model is its omission of baroclinic instability, for in fast rotating systems baroclinic dynamics are likely to play a significant role in heat redistribution. The one-layer model of itself also does not exhibit the superrotation seen the in 1.5-layer shallow-water model of \cite{Showman:2010hk}, and advective transport by a superrotating jet may be a significant effect. Studies of this, and the use of a stratified three-dimensional model, are topics we hope to present in a future publication.
